\documentclass[%
 reprint,
superscriptaddress,
preprintnumbers,
nofootinbib,
 amsmath,amssymb,
 aps,
]{revtex4-2}

\usepackage{graphicx}
\usepackage{dcolumn}
\usepackage{bm}
\usepackage{hyperref}

\usepackage[usenames,dvipsnames]{xcolor}
\usepackage{hyperref}
\hypersetup{
    colorlinks = true,
    citecolor = {magenta},
    linkcolor = {blue},
    urlcolor = {blue}		
}
\usepackage{mathtools}
\usepackage{aas_macros}

\begin{document}

\title{
Significance of void shape: Neutrino mass from Voronoi void halos?
}

\author{Adrian E.~Bayer}
\email{abayer@princeton.edu}
\affiliation{
Department of Astrophysical Sciences, Princeton University, Peyton Hall, Princeton NJ 08544, USA
}%
\affiliation{
Center for Computational Astrophysics, Flatiron Institute, 162 5th Avenue, New York NY 10010, USA
}%

\author{Jia Liu}
\affiliation{
Center for Data-Driven Discovery, Kavli IPMU (WPI), UTIAS, The University of Tokyo, Kashiwa, Chiba 277-8583, Japan
}

\author{Christina D.~Kreisch}
\affiliation{
Department of Astrophysical Sciences, Princeton University, Peyton Hall, Princeton NJ 08544, USA
}%

\author{Alice Pisani}
\affiliation{
Aix-Marseille University, CNRS/IN2P3, CPPM, 163 Av. de Luminy, 13009, Marseille, France}

\affiliation{
Department of Astrophysical Sciences, Princeton University, Peyton Hall, Princeton NJ 08544, USA
}%

\affiliation{
Center for Computational Astrophysics, Flatiron Institute, 162 5th Avenue, New York NY 10010, USA
}%
\affiliation{
The Cooper Union, 41 Cooper Square, New York NY 10003, USA
}%

\date{\today}

\begin{abstract}
Massive neutrinos suppress the growth of cosmic structure on nonlinear scales, motivating the use of information beyond the power spectrum to tighten constraints on the neutrino mass, for example by considering cosmic voids.
It was recently proposed that constraints on neutrino mass from the halo mass function (HMF) can be improved by considering only the halos that reside within voids -- the void-halo mass function (VHMF). 
We extend this analysis, which made spherical assumptions about the shape of voids, to take into account the non-spherical nature of voids as defined by the Voronoi-tessellation-based void finder, \texttt{VIDE}.
In turn, after accounting for one spurious non-spherical void, we find no evidence that
the VHMF contains information beyond the HMF.
Given this finding, we then introduce a novel summary statistic by splitting halos according to the emptiness of their individual environments, defined by the Voronoi cell volume each halo resides in, and combining the mass functions from each split. 
We name the corresponding statistic the \textit{VorHMF} and find that it could provide information regarding neutrino mass beyond the HMF. 
Our work thus motivates the importance of accounting for the full shape of voids in future analyses, both in terms of removing outliers to achieve robust results and as an additional source of cosmological information.
\end{abstract}

\maketitle


\section{Introduction}
\label{sec:intro}

Upcoming cosmological surveys such as DESI 
\citep{collaboration2016desi}, PFS 
\citep{Takada_2014}, the Rubin Observatory LSST 
\cite{LSSTSci},
Euclid 
\cite{Euclid}, 
SPHEREx 
\cite{SphereX_2014}, 
SKA 
\cite{SKA_2009}, and the Roman Space Telescope 
\cite{spergel2013widefield} will probe decreasingly small scales of cosmic structure, where massive neutrinos suppress the growth of cosmic structure. 
By computing the nonlinear effects of massive neutrinos on structure formation 
\cite{Liu2018MassiveNuS:Simulations, Bird2018, Villaescusa_Navarro_2018, Arka_2018, Bayer_2021_fastpm, DeRose:2023dmk, Sullivan:2023ntz},
it has been argued that
the underdense regions of the cosmic web, known as voids, will be particularly sensitive to the sum of the neutrino masses $M_\nu$ \cite{Massara2015,Kreisch2019,Schuster2019,contarini2022,verza2023}. It was shown in \citep{bayer2021detecting} that voids in the matter field contain a large amount of information that cannot be found in the power spectrum or halo mass function. These findings were further investigated in the context of the halo field in \cite{Kreisch_2021}, and, in particular, \cite{Zhang_2020} showed a significant $M_\nu$ signal when considering the mass function of halos residing within voids. This statistic, known as the void-halo mass function (VHMF), was shown to be significantly more constraining than the halo mass function (HMF) computed using all halos, however, the analysis relied on spherical assumptions for the void-halo finding.

In this paper, we first revisit the analysis of \cite{Zhang_2020} while accounting for asphericity. We find that, after removing a spurious and anomalously non-spherical void from the analysis, the VHMF contains no more information than the HMF for the minimum halo mass  considered ($M_h\geq3\times10^{11}\,M_\odot/h$). We then move beyond spherical assumptions by considering halos residing in the full void shape defined by \texttt{VIDE} \citep{VIDE}, again finding that the VHMF contains no more information than the HMF. This lack of signal is due to \texttt{VIDE} including dense ``void walls'' as part of its void definition. Thus, as a first attempt to exclude void walls from the analysis, we then consider the mass function of each individual Voronoi cell and split the mass function based on the Voronoi cell volume -- we call this the VorHMF. We find evidence that this provides complementary information to the HMF, making it a 
fruitful summary statistic to study in more depth.

The structure of this paper is as follows. In Section \ref{sec:analysis} we outline the methodology, first reviewing the simulations used in this work, followed by the halo mass function and void finding procedure. In Section \ref{sec:results} we quantify the constraining power of the various void and Voronoi-based mass functions in comparison to the HMF.
Finally, in Section \ref{sec:conclusion} we conclude and discuss our results, in particular commenting on the importance of accounting for the full shape of voids, both in terms of cutting out outliers and as an additional source of information to constrain cosmology.

\section{Methodology}
\label{sec:analysis}

We first reproduce the analysis of \cite{Zhang_2020}, for which we now outline the methodology.

We use the \texttt{MassiveNuS} simulations, which are based on a modified version of the N-body code \texttt{Gadget-2}~\citep{Gadget}, where neutrinos are treated as linear perturbations to the matter density, but their evolution is sourced by the full nonlinear matter perturbations. 
We use the two fiducial, seed-matched, simulations with  $M_\nu=0.1\rm{eV}$ and $0.6\rm{eV}$.  All other cosmological parameters are fixed at $A_s = 2.1 \times 10^{-9}$, $\Omega_m=0.3$, $\Omega_b = 0.05$, and $h = 0.7$.
Each simulation contains $1024^3$~particles in a  (512Mpc$/h$)$^3$ box.

We use the publicly available halo catalogs for \texttt{MassiveNuS} which employ the \texttt{ROCKSTAR} halo finder \citep{ROCKSTAR}.
The HMF, denoted $n_h$, is defined as the comoving number density of halos per unit of log (base 10) halo mass.
We consider 9 logarithmic bins bounded by $3\times 10^{11} M_\odot/h$ and $3 \times 10^{15} M_\odot/h$. 
We consider redshift $z=0$ 
where the signal is expected to be the strongest.


We find voids using the public \texttt{VIDE} algorithm\footnote{\url{https://bitbucket.org/cosmicvoids/vide_public/wiki}} \citep{VIDE} (based on the earlier \texttt{ZOBOV} algorithm \citep{neyrinck2008}). We consider voids defined in the halo field, rather than the matter field, as this is closer to what is observed in reality. \texttt{VIDE} first  performs Voronoi tessellation in which each halo is associated with a Voronoi cell. Each cell has an associated volume, with larger volume implying larger isolation from other halos (i.e.~a halo belonging to a more void-like region). \texttt{VIDE} then performs a watershed transform to group these Voronoi cells into larger structures referred to as voids. While this defines the full, non-spherical, shape of voids, we also define the \textit{effective radius} of a void ($R_v$) as the radius of a sphere with an equivalent volume. We note that \texttt{VIDE} is applied to all halos in the MassiveNuS catalog while a minimum mass cut of $3\times 10^{11} M_\odot/h$ is applied when computing mass functions, following the analysis of \cite{Zhang_2020}; investigating the effects of mass cuts would be fruitful future work, but for the purposes of this paper we mimic the previous analysis.

Due to a number of effects, the void finding procedure can provide voids that are deemed to be unphysical, for example, the presence of shot noise causes the algorithm to output spurious shallow voids \citep{neyrinck2008,pisani2015b,nadathur2015natureI,nadathur2015natureII}. The void catalog is thus typically cleaned to remove spurious voids in a post-processing step. In this work we mimic the analysis of \cite{Zhang_2020}, which uses the cleaning procedure of the default \texttt{VIDE} catalog \cite{VIDE, sutter2012public}, which only considers parent voids and applies a central density cut of $\rho (R<0.25R_v) > 0.2\bar{\rho}$, where $\rho(R)$ is the density at radius $R$ from the void center, and $\bar{\rho}$ is the mean density. 
We also tested the effect of relaxing this cut, 
finding no qualitative difference to our conclusions. 


To define the VHMF, the methodology of \cite{Zhang_2020} was to locate halos within a sphere centered on the void center, with radius equal to half the void's effective radius ($0.5 R_v$). The motivation for this is that halos far away from the void center -- near void walls -- may experience complicated galaxy formation processes and should not be treated as void-halos. However, this analysis assumes voids to have a somewhat spherical nature, while still defining voids using a non-spherical void finder. In this work we consider the full void shape output by \texttt{VIDE}, identifying a void-halo as any halo within the \texttt{VIDE}-defined full shape of a void. This is the natural extension of the VHMF definition used in \cite{Zhang_2020}, however, since \texttt{VIDE} voids include the walls of voids -- the over-dense shells of voids -- it is expected that this will include halos that are in locally over-dense environments and thus this is not the most fruitful definition to probe information beyond the HMF. We then consider the individual Voronoi cells from the Voronoi tessellation, computing the mass function for halos with different Voronoi cell volume cuts -- the VorHMF. Splitting the halo mass function according to the Voronoi cell volume, and combining the different cuts, folds in information about the local density of each halo.
To make a fair comparison, we normalize the mass functions by the volume of space in which voids are sought: for the HMF this is the volume of the entire box, for the $0.5 R_v$VHMF this is the sum of the half-sphere volumes for each void, for the full-shape VHMF this is the sum of the full-shape volumes for each void, and for the VorHMF this is the sum of the Voronoi cell volumes. This is the same normalization choice as \cite{Zhang_2020} and does not affect the qualitative results.

\section{Results}
\label{sec:results}

\begin{figure*}[!t]
\includegraphics[width=0.49\linewidth]{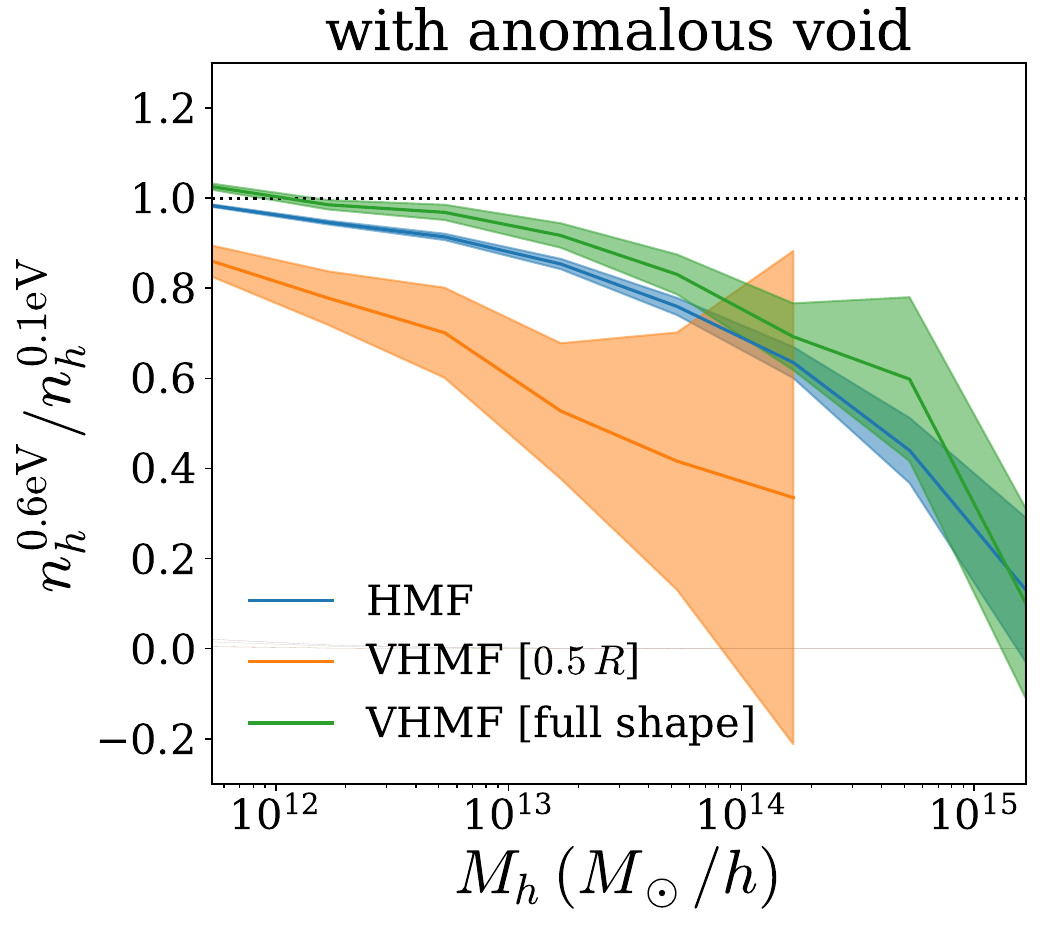}
\includegraphics[width=0.49\linewidth]{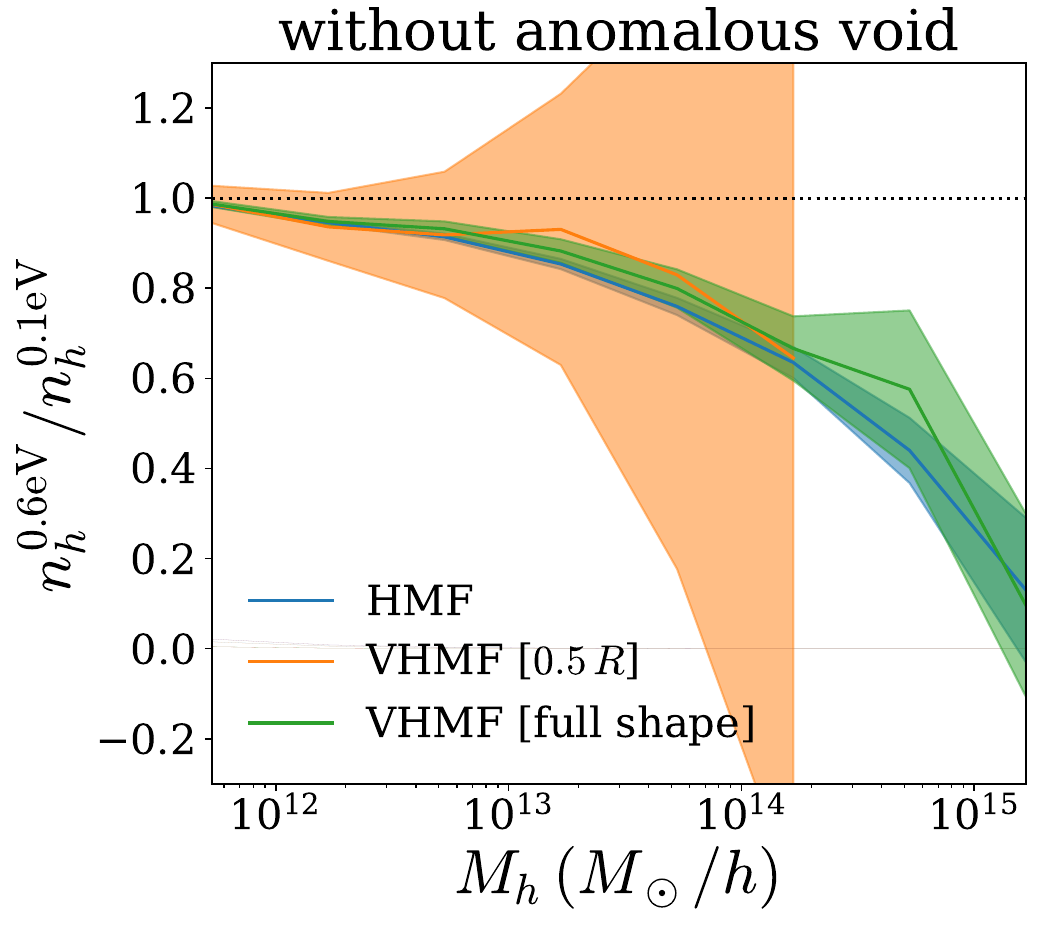}
\caption{The ratio between a cosmology with $M_\nu = 0.6 {\rm eV}$ and $0.1 {\rm eV}$ for the HMF (blue), spherical $0.5 R_v$VHMF (orange), and full-shape VHMF (green). Error bars correspond to 95\% confidence intervals. The \textit{left} plot includes the one spurious void, in turn giving a seemingly large signal for the spherical VHMF, and also a discrepancy between the full-shape VHMF and the HMF. The \textit{right} plot repeats the analysis without the spurious void, showing that both definitions of the VHMF are consistent with the HMF, implying that there is no additional information regarding neutrino mass in the VHMF that is not in the HMF.
}
\label{fig:ratio} 
\end{figure*}

We first consider
the VHMF defined using halos within spheres with radius equal to half the effective void radius ($0.5 R_v$) centered on the void center. Fig.~\ref{fig:ratio} considers the ratio of the HMF and VHMF between a cosmology with $M_\nu=0.6{\rm eV}$ and $0.1{\rm eV}$. Error bars correspond to the 95\% confidence intervals and are approximated using Poisson statistics.
The left panel compares the $0.5 R_v$VHMF (orange) to the total HMF (blue).  As found in \cite{Zhang_2020}, the $0.5 R_v$VHMF ratio is significantly different to the HMF. For example, while the HMF ratio is unity for small halo masses of $10^{12} M_\odot/h$, there is an approximately 20\% difference from unity for the $0.5 R_v$VHMF. This suggests that neutrino mass causes a significantly larger effect on the VHMF compared to the HMF, and that halos in voids are particularly sensitive to neutrino mass.

However, when taking into account the full shape of voids (green) it can be seen that the ratio for the VHMF becomes similar to the HMF. This is predominantly because voids are not spherical, and thus the $0.5 R_v$VHMF includes some halos which are not void-halos. There is however still some difference between the the HMF and full-shape VHMF which we will now explain.


\begin{figure}[!t]
\includegraphics[width=\linewidth]{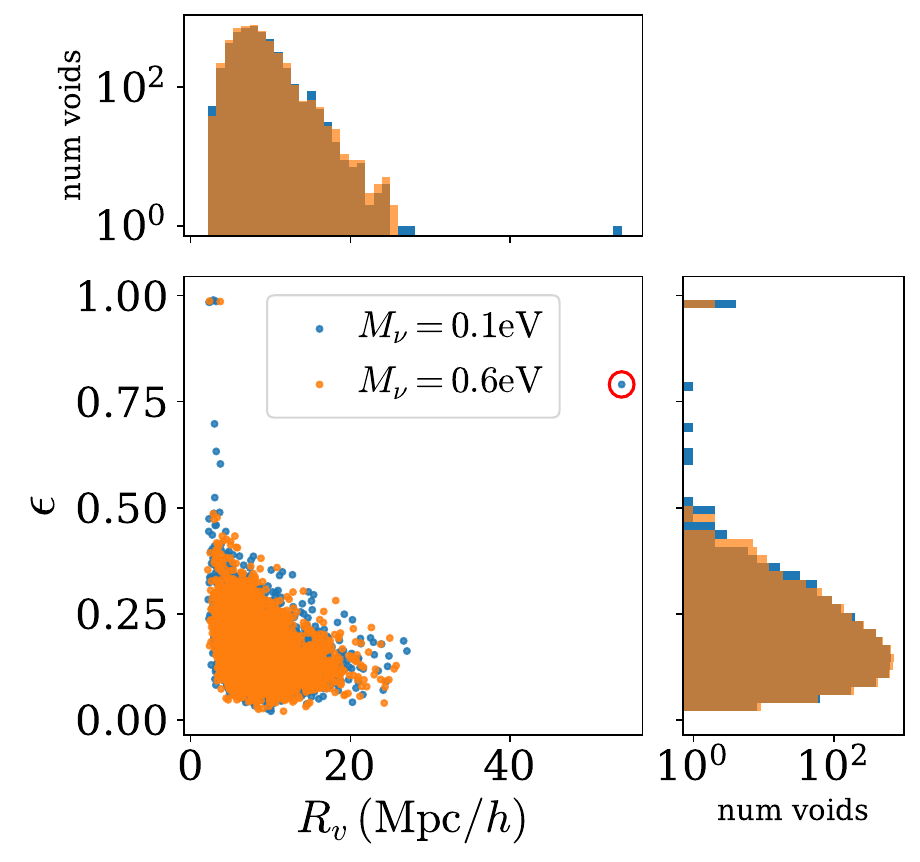}
\caption{A scatter plot (center) of void ellipticity versus effective radius, and their respective histograms (top and right), for cosmologies with $M_\nu = 0.1 {\rm eV}$ (blue) and $0.6 {\rm eV}$ (orange). A prominent outlier can be seen in the $M_\nu = 0.1 {\rm eV}$ cosmology at $R_v=54{\rm Mpc}/h, \epsilon=0.8$ (circled in red). There are also some highly elliptical voids with small $R$ -- this is because for a void with small volume, a minor perturbation to the length of one of its axes can cause a high level of ellipticity -- but we find that removing these voids does not modify the conclusions of the VHMF in this work.
}
\label{fig:radvellip} 
\end{figure}

\begin{figure}[!t]
\includegraphics[width=\linewidth]{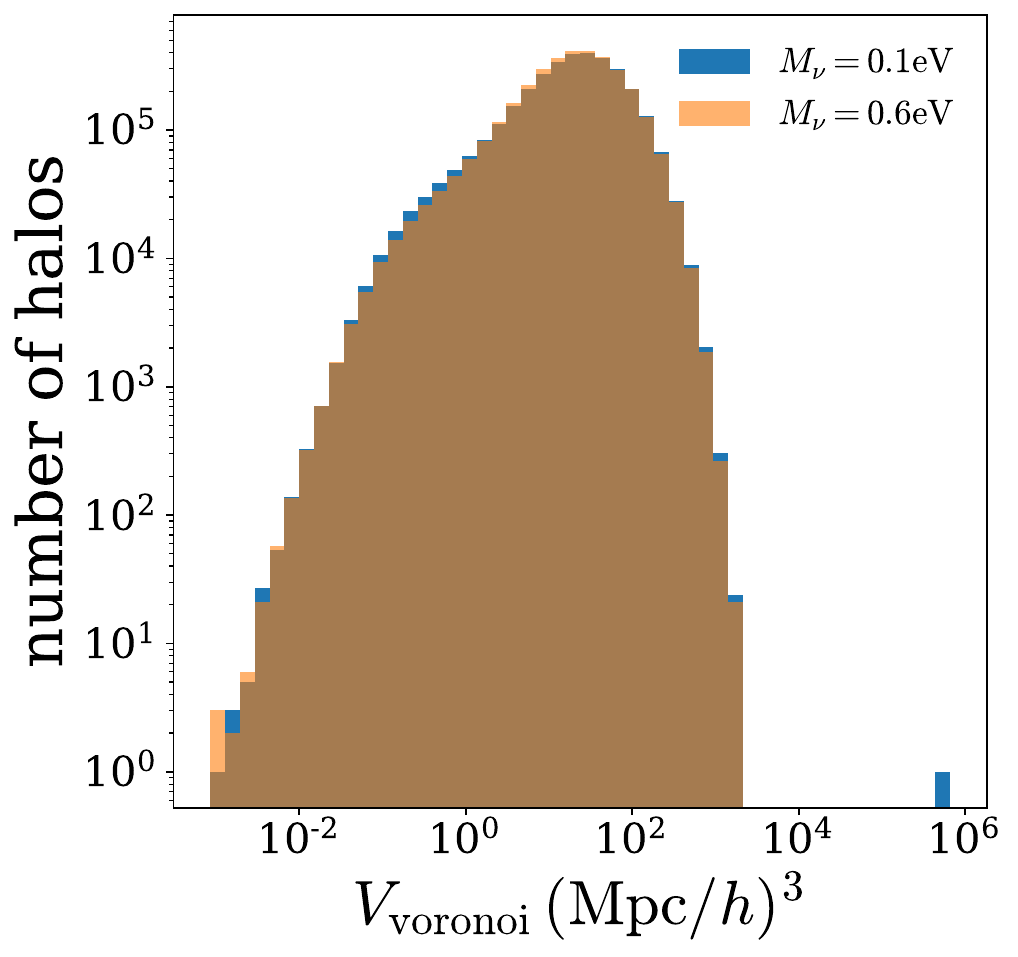}
\caption{A histogram of Voronoi cell volumes in the \texttt{MassiveNuS} simulation for cosmologies with $M_\nu = 0.1 {\rm eV}$ (blue) and $0.6 {\rm eV}$ (orange). We also show the spurious Voronoi cell.}
\label{fig:hist_vor} 
\end{figure}

\begin{figure}[!t]
\includegraphics[width=\linewidth]{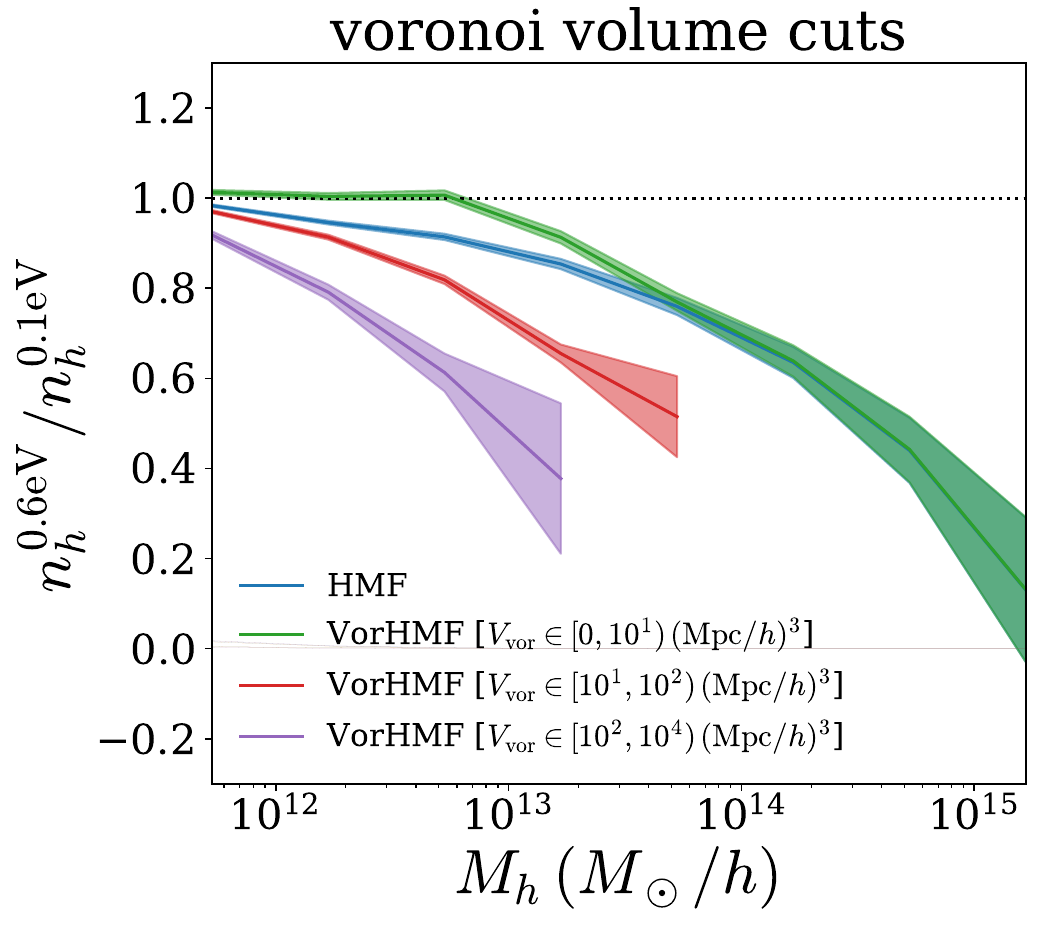}
\caption{Voronoi cell analysis for \texttt{MassiveNuS}. The ratio between a cosmology with $M_\nu = 0.6 {\rm eV}$ and $0.1 {\rm eV}$ for the HMF (blue), and the VorHMF splits for various ranges of Voronoi volume cuts. The ratio between neutrino mass cosmologies diverges from unity more as larger $V_{\rm vor}$ is considered. This implies that splitting halos according to the Voronoi volume in which they reside, and combining the mass functions from each split, could provide more information regarding neutrino mass than the HMF with no splitting. 
}
\label{fig:vor} 
\end{figure}

\begin{figure}[!t]
\includegraphics[width=\linewidth]{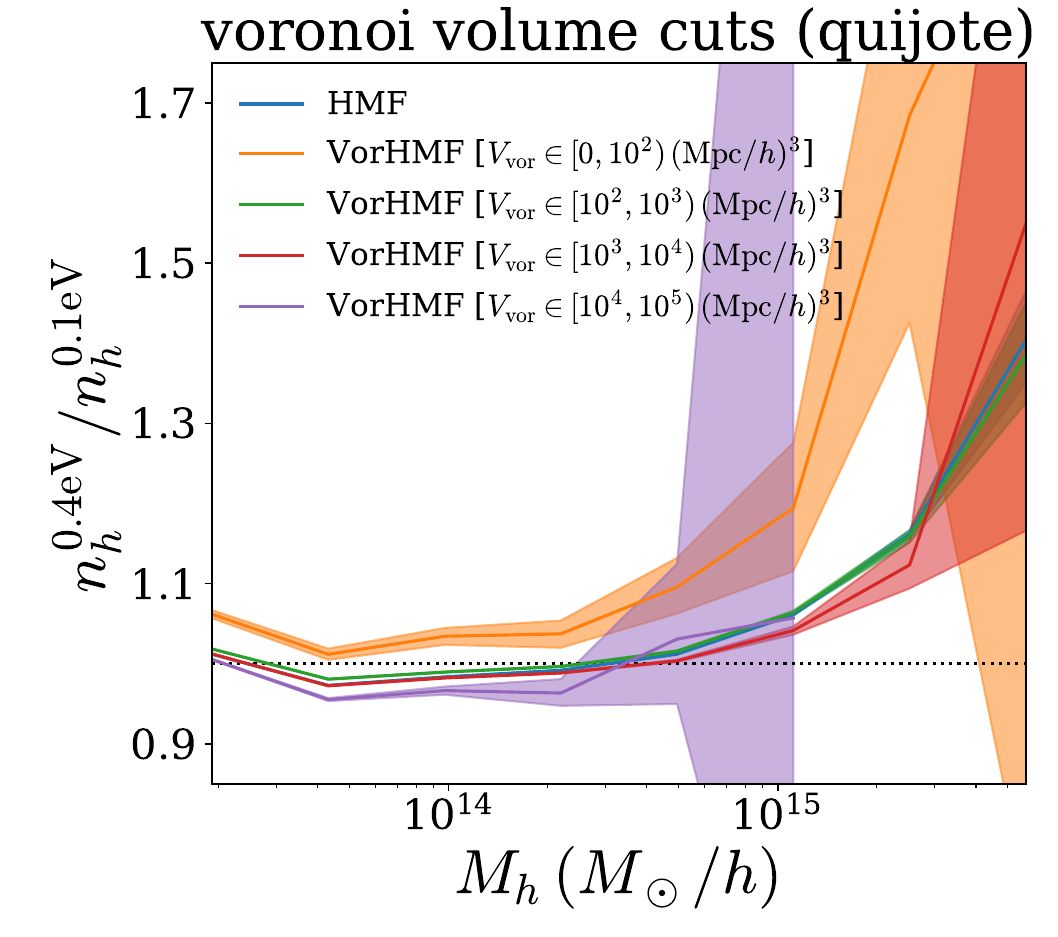}
\caption{Voronoi cell analysis for the \texttt{Quijote} simulations. The ratio between a cosmology with $M_\nu = 0.4 {\rm eV}$ and $0.1 {\rm eV}$ for the HMF (blue), and the VorHMF splits for various minimum Voronoi volume cuts.
}
\label{fig:quijote} 
\end{figure}


The top panel of Fig.~\ref{fig:radvellip} shows a histogram of the voids as a function of their radius. We find one void with a particularly large radius of $54\,{\rm Mpc}/h$ exists for the $0.1{\rm eV}$ cosmology, but not for $0.6{\rm eV}$. This lack of one-to-one correspondence between voids in the two cosmologies distorts the ratio. Moreover, not only is this void an outlier in terms of radius, but it is also highly non-spherical: Fig.~\ref{fig:radvellip} also shows that this void has an ellipticity of 0.8.
Furthermore, we find that this void contains 5 void-halos out of the 588,509 void-halos found in the full shape analysis, whereas it contains 2,817 of the 17,371 void-halos in the $0.5 R_v$ analysis. This means that this one void (out of 4,820 voids) is responsible for 16\% of the void-halos in the $0.5 R_v$VHMF, and that essentially all of these halos are not actually void-halos.
The volume of this void also contributes 3.8\% of the total void volume, which further distorts the ratio. 

All of the above 
suggests that this void should be removed from the analysis. 
The right panel of Fig.~\ref{fig:ratio} shows the same plot as the left panel, but with the one anomalous void removed from the analysis. Now all (V)HMF lines agree, suggesting there is no additional information regarding neutrino mass in the VHMF that cannot be found in the HMF, moreover, the scatter in the VHMF is much larger.
It is also interesting to note that the full shape VHMF is far less sensitive to the anomalous void compared to the $0.5 R_v$VHMF -- this can be seen from the similarity (dissimilarity) of the green (orange) lines in the left and right plot -- indicating that a spherical selection increases the impact of spurious voids on the results.
We additionally studied a seed-matched simulation with $M_\nu=0$ and found no anomalous voids, implying that the anomaly is somewhat unique to the $0.1\,{\rm eV}$ simulation.

Given these findings, we now ask if neutrino mass information could be found by looking at the smaller-scale underlying structures used to define voids. In \texttt{VIDE}, each halo belongs to a Voronoi cell. Voronoi cells are then grouped together using a watershed transform to form voids, but one could drop this watershed transform and consider each individual Voronoi cell without grouping them together. 
Each Voronoi cell has a volume $V_{\rm vor}$; a higher $V_{\rm vor}$ implies a more isolated halo and thus a halo belonging to a more underdense region. 
Fig.~\ref{fig:hist_vor} shows histograms of $V_{\rm vor}$, including the one spurious Voronoi cell which was responsible for the spurious void in the previous analysis. 

To study the relationship between $V_{\rm vor}$ and the mass function, Fig.~\ref{fig:vor} shows the mass function computed for halos within different bins of $V_{\rm vor}$. It can be seen that the larger the Voronoi cells, the larger the ratio of the mass function between the different neutrino mass cosmologies, implying that halos that are most isolated have the strongest response to neutrino mass. 
This implies that by performing a combined analysis on the mass functions computed within different $V_{\rm vor}$ bins -- the VorHMF -- one could obtain information beyond the HMF. This is similar logic to density-split approaches \cite{Paillas:2022wob}. It is also interesting to note that while we excluded the spurious Voronoi cell (in Fig.~\ref{fig:hist_vor}) from the analysis, the results of Fig.~\ref{fig:vor} look identical even if it were included; this is because the spurious Voronoi cell only constitutes $0.5\%$ of the volume of the box, while the spurious void constitutes $3.8\%$ of the void volume
, making the effect of a spurious Voronoi cell weaker than a spurious void. Hence, this approach is more robust to non-spherical outliers, as well as providing more information.


One key consideration when determining if a higher-order statistic is sensitive to neutrino mass is whether it is able to break the $M_\nu-\sigma_8$ degeneracy \cite{bayer2021detecting}. In Fig.~\ref{fig:quijote} we repeat the VorHMF analysis using the \texttt{GIGANTES} \cite{Kreisch_2021} catalog from the \texttt{Quijote} simulations \cite{quijote}. We use the 500 $M_\nu^+$ and $M_\nu^{+++}$ simulations for this analysis, corresponding to neutrino masses of $0.1\,{\rm eV}$ and $0.4\,{\rm eV}$ respectively. Note there are three key differences between \texttt{MassiveNuS} and \texttt{Quijote}: (i) \texttt{MassiveNuS} has fixed $A_s$, while \texttt{Quijote} has fixed $\sigma_8$, (ii) the minimum halos mass of \texttt{MassiveNuS} is lower than \texttt{Quijote}, and (iii) the maximum neutrino mass in \texttt{Quijote} is $0.4\,{\rm eV}$ instead of $0.6\,{\rm eV}$ in \texttt{MassiveNuS}.
It can be seen that there is a modest signal in \texttt{Quijote}, in particular the low $V_{\rm vor}$ bin deviates from the HMF the most. The effect is smaller than in \texttt{MassiveNuS}, either because (i) the VorHMF constraints on neutrino mass are somewhat degenerate with $\sigma_8$ and thus another probe will be required to break this degeneracy (such as the CMB or other higher-order statistics \cite{bayer2021detecting})); and/or (ii) the VorHMF is not sensitive to neutrino mass at the higher minimum halo mass of the \texttt{Quijote} simulations; and/or (iii) \texttt{Quijote} has a lower maximum neutrino mass.
We also note that the algorithms used to simulate neutrinos are different in each simulation -- \texttt{MassiveNuS} uses the linear response approach \cite{Ali_Ha_moud_2012}, while \texttt{Quijote} uses neutrino particles -- but both of these methods have been shown to converge equivalently so we expect this to have little effect \cite{Ali_Ha_moud_2012, Bird2018}.
It is also possible that this is due to spurious voids or resolution effects --
it would thus be fruitful further work to study the VorHMF with a new set of high resolution simulations to quantify exactly how much information it contains as a function of minimum mass cut and Voronoi volume.

\section{Discussion and Conclusions}
\label{sec:conclusion}
We revisited the neutrino mass information content of the VHMF, first analyzed in \cite{Zhang_2020} with spherical assumptions about the void interior. We showed that, after removing one spurious non-spherical void, there is no information in the VHMF that cannot also be found in the HMF. This is true both when making spherical assumptions or when considering the full shape of voids defined by \texttt{VIDE}, but considering the full shape greatly reduces the impact of spurious voids. Finally, we showed that considering the mass function of each individual Voronoi cell, and splitting based on Voronoi cell volume, has the potential to provide information beyond the HMF. We named this novel summary statistic the VorHMF. Thus we have shown that considering the full shape of voids is important both in terms of removing outliers to achieve robust results and as a potential additional source of cosmological information.


While the VorHMF has been shown to contain information beyond the HMF in principle, the next step would be to investigate how much information it contains in practice. It would be instructive to study on what scales it contains complementary information to the halo power spectrum, à la \cite{Bayer:2021kwg}, and also for what halo mass resolutions.  Moreover, it will be fruitful future work to perform a full parameter inference to see how much information remains after accounting for the degeneracies between $M_\nu$ and other parameters, in particular $\sigma_8$ \cite{bayer2021detecting} and bias parameters, and after taking into account cosmic variance. Furthermore, it will be important to explore how robust the VorHMF is to small-scale systematics, noise, and baryonic effects. It will also be interesting future work to perform higher resolution simulations to study the VHMF and VorHMF for lower mass halos, where the effect of neutrinos will be more prominent. Additionally, one could consider different properties and cuts when defining the VorHMF -- for example, one could consider a VorHMF using only Voronoi cells that are within a given distance of the void center and/or cells above a certain Voronoi volume. Work towards understanding how other cosmological properties vary with Voronoi cell volume was made in \cite{Jamieson:2020qkn}. All of the above, paired to work on theoretical modeling, will pave the way to applications of this methodology to data from current and upcoming surveys.

While this work only considered the information gain from splitting the HMF by the Voronoi volume, one could also consider other higher order statistics computed when splitting halos according to their Voronoi volume. One could also consider including the Voronoi volume as an input parameter for machine learning approaches, for example, in graph neural network approaches where each halo corresponds to a node in the graph \cite{Villanueva-Domingo:2021dun, Makinen:2022jsc, Wu:2023exg, deSanti:2023zzn, wang2023} one could include the Voronoi volume as a node property to encode void-like information.


In addition to highlighting the importance of considering void shape, this work has illustrated spurious signals can arise from the lack of one-to-one correspondence between voids in different cosmologies. 
In this work we found that one spurious void existed in the $M_\nu=0.1{\rm eV}$ cosmology which did not exist in the $M_\nu=0.6{\rm eV}$ cosmology, and this lead to an unphysical conclusion in \cite{Zhang_2020}. In principle, spurious signals could arise in any analysis that compares objects between a set of simulations with a discrete change of cosmological parameters, for example when (i) taking the ratio between simulations with two different sets of parameters, (ii) using finite difference methods for derivatives, or (iii) performing simulation-based inference using a hypercube of simulations with different parameters. Care must thus be taken to use a sufficiently small step size between parameters and to run sufficiently many simulations, or sufficiently large simulations, to ensure a sufficiently high number of voids and reduce the effects of such outliers. 

It is important to consider the interpretation of the spurious void found in this analysis. Is a highly non-spherical void something that should be considered physical and modelled when analyzing data? 
Does the presence of such voids have a smooth dependence on the cosmology?
Or should such a void be considered unphysical and removed from the analysis?
Much previous literature has considered these voids to be unphysical, for example, the presence of shot noise has been noted to cause void finders to output spurious shallow voids \citep{neyrinck2008,pisani2015b,nadathur2015natureI,nadathur2015natureII}. The void catalog is thus typically cleaned to remove spurious voids in a post-processing step, and various cleaning methods have been proposed \citep{ronconi2017,contarini2022}.
For example, cuts can be made based on the density contrast \citep{sutter2012public, Mao_2017}, the central density \citep{nadathur2015natureI}, the central shape of the density profile \citep{Sutter_2014}, the void hierarchy \cite{VIDE}, and the mean particle separation \citep{pisani2015b,Hamaus_2016,Hamaus_2017, contarini2022,contarini2023,Thiele:2023oqf}. 
The lack of objective cleaning procedure poses a problem in terms of having consistency between different void analyses; to work towards minimizing this, multivariate machine learning approaches have been developed to take into account all void properties in unison \citep{Cousinou:2018lhi}. 
A general solution may not exist, with more stringent cleaning procedures potentially becoming important when analyzing the interior structure of voids, such as in this work, or galaxy properties in voids, in particular in the case of a small overall volume leading to low void numbers \citep{habouzit2020}.
Our work has further motivated the use of considering shape information in void analyses. In the case of \citep{Zhang_2020}, a simple maximum radius or ellipticity cut would have been sufficient. However, in general, the presence of spurious objects could be subtle, motivating the importance of developing robust methods to remove all spurious objects from catalogs. These problems could also arise in the context of halo finding, for which many algorithms also rely on post-process cleaning procedures. In fact, it's possible that the spurious void in this analysis was the product of spurious halos, as the void catalog is computed using the halo catalog.
It would thus be fruitful future work to test the practical importance of these findings by comparing the inference of cosmological parameters obtained when using different cleaning procedures.
On the other hand, the anomalous void may in fact hint at a physical effect, with massive neutrinos tending to make voids less elliptical. It would thus be fruitful future work to study the information content of void shape in more depth \citep{lavaux2010,Bos_2012,wang2023}.\\

\break
\begin{acknowledgments}
We thank Francisco Villaescusa-Navarro, Benjamin Wandelt, and Gemma Zhang for insightful comments and discussion.
This work took place in 2020/1 when AEB and JL were at the University of California, Berkeley. 
This work was supported by JSPS KAKENHI Grants 23K13095 and 23H00107 and the MEXT Program for Promoting Researches on the Supercomputer Fugaku hp230202 (to JL). 
AP acknowledge support from the Simons Foundation to the Center for Computational Astrophysics at the Flatiron Institute and support from the European Research Council (ERC) under the European Union's Horizon programme (COSMOBEST ERC funded project, grant agreement 101078174). 
This work used Princeton Research Computing resources at Princeton University which is a consortium of groups led by the Princeton Institute for Computational Science and Engineering (PICSciE) and Office of Information Technology's Research Computing.

\end{acknowledgments}

\bibliography{references}

\end{document}